\documentclass{iaus}
\usepackage{graphicx}
\usepackage{subfigure}
\def\ltsima{$\; \buildrel < \over \sim \;$}
\def\simlt{\lower.5ex\hbox{\ltsima}}
\def\gtsima{$\; \buildrel > \over \sim \;$}
\def\simgt{\lower.5ex\hbox{\gtsima}}

\title[Dark Matter on Small Scales] %% give here short title %%
{Observed Properties of Dark Matter on Small Spatial Scales}

\author[Wyse and Gilmore]   %% give here short author list %%
{Rosemary F.G.~Wyse$^1$%
\break \and  Gerard Gilmore$^2$}

\affiliation{$^1$Department of Physics \& Astronomy, Johns Hopkins University, Baltimore, MD 21218, USA  \break email: wyse@pha.jhu.edu\\[\affilskip]
$^2$Institute of Astronomy, 
Cambridge, CB3 0HA,  UK \break email: gil@ast.cam.ac.uk}

\pubyear{2007}
\volume{244}  %% insert here IAU Symposium No.
\pagerange{}
\date{?? and in revised form ??}
\jname{Dark Galaxies \& Lost Baryons}
\editors{J. Davies, ed.}
\begin{document}

\maketitle

\begin{abstract}
The nature of dark matter is one of the outstanding questions of
astrophysics. The internal motions of member stars reveal that the
lowest luminosity galaxies in the Local Group are the most dark-matter
dominated. New large datasets allow one to go further, and determine
systematic properties of their dark matter haloes. We summarise recent
results, emphasising the critical role of the dwarf spheroidal
galaxies in understanding both dark matter and baryonic processes that
shape galaxy evolution.  
\end{abstract}

\section{Introduction}

The study of dark matter is best undertaken in systems that are the
most dark-matter dominated, and for which the baryon content has had
minimal effect on the dark matter halo.  The dwarf spheroidal
satellite galaxies of the Milky Way are such systems, with their high total 
dark-matter content being inferred some 25 years ago (\cite{Aaronson83})
from their central stellar velocity dispersions (based on as few as
three stars). Modern multi-object spectrographs on large telescopes
make it possible now to acquire and analyse statistically significant
samples of stars (many hundreds) across the face of these systems. This allows derivation of mass profiles. 

Study of these smallest systems has the further advantage that it is
on the smallest scales where the predictions of galaxy formation
models with different types of dark matter diverge and are most easily
discriminated (\cite{Ost03}).

The dwarf spheroidal galaxies (dSph) are low surface-brightness,
gas-poor systems, identified through star counts.  They are the most
common galaxy in the local Universe.  The \lq classical' dSph (see
e.g.~\cite{jsg94}; \cite{mateo98}) were generally detected prior to
1990, are at distances of $\sim 70$~kpc to $\sim 150$~kpc, have
typical total luminosity of $\sim 10^7$L$_{V,\odot}$, characteristic
surface brightness of $\mu_V \sim 24$~mag/sq~arcsec and are extremely
gas-poor. The central stellar velocity dispersion is $\sim 10$~km/s,
and combined with the characteristic radius of a few hundred parsec, 
this implies mass-to-light ratios in solar units in the range of $ 10
\simlt M/L_V \simlt 300$. The stars are metal-poor, with a typical mean
[Fe/H] $\simlt -1.5$~dex. All dSph contain old stars, and initiated
star formation at early times, corresponding to a lookback time of
$\simgt 12$~Gyr. More surprisingly, given the expectation of early
supernovae-driven winds from these shallow potential wells
(e.g.~\cite{dekel86}; see also \cite{wyse85} for a simpler derivation
of the threshold escape velocity) most dSph contain stars of a very
broad range of ages, and indeed the average dSph member star is of
intermediate-age (e.g.~\cite{smecker94}; \cite{xavier00}). If there
were (supernovae-driven?) outflows, the gas must have been
re-accreted.\footnote{While outwith the scope of this talk, there is a real 
lack of
cosmologically consistent models of chemical evolution of the dSph
that also are consistent with the individual star formation histories and do not invoke {\it ad hoc\/} outflows and inflows; see \cite{sws87} for an early attempt.}

The dSph are then among the first systems to collapse and form stars,
and have potential-well depths shallow enough for the baryonic content
to be considerably affected by reionization of the Universe (a one-dimensional 
velocity dispersion of 10km/s
corresponds to of order $\sim 10^4$~K, approximately the ionization equilibrium temperature of hydrogen; see \cite{gpe92}; \cite{bkw00}). This potential-well depth is also sufficiently
shallow that, as noted above, internal effects -- such as ionizing photons from massive stars, and energy and
momentum injection from stellar winds and supernovae -- can have a significant effect.  Ram pressure, as the dwarf
galaxy moves through the intergalactic medium (or through the gaseous
halo of the Milky Way), can also cause gas to be removed from the
dwarf.  Thus the star formation histories and chemical evolutions of
the dSph, which can be derived from observations such as deep
colour-magnitude diagrams combined with spectroscopic elemental abundance
measurements, can be used to constrain the physics assumed in theories
of galaxy evolution that invoke significant \lq feedback' on galactic
scales to modify their stellar content e.g.~\cite{springel05}. \lq
Feedback' is not a free parameter.

As noted, these are the most dark-matter dominated galaxies. Analyses
of large samples of internal velocities in galaxies over as wide a
range of luminosity as possible will allow the quantification of
trends in the inferred properties of the dark matter content with, 
e.g., galaxy luminosity or scale-length, and thereby place constraints
on the type of dark matter (e.g.~\cite{dekel86}; \cite{kf04};
\cite{zar06}). Whether or not the dSph fit smoothly onto
extrapolations of scaling relations seen for normal large galaxies is
an interesting question, with implications for dark matter. The
(model-dependent, at present) mass profile of individual dark haloes
may be obtained from the stellar line-of-sight velocities, and
compared to the predictions for different types of dark matter.  It has
long been known that the simplest variant of $\Lambda$CDM predicts too
many dwarf-galaxy-mass dark haloes compared to the number of satellite
galaxies (\cite{moore99}; \cite{klypin99}) and the derived mass functions and luminosity functions of
the observed satellites are crucial in testing models, as are their
detailed stellar populations.

The sample of known dSph satellites of the Milky Way has been recently
increased by a factor of about two, primarily through the analysis of
the uniform wide-field photometry of the stellar sky from the Sloan
Digital Sky Survey.  These new detections are of significantly lower
luminosity, extending down to $\simlt 10^3$L$_\odot$, and lower mean
surface brightness, typically $\mu_V \sim 30$~mag/sq~arcsec (see
\cite{field06}, \cite{blobs07} and references therein.)  Thus these
galaxies overlap with star clusters in terms of total luminosity, and
the question of what distinguishes star clusters from galaxies gains
new relevance. Apart from the existence of a dark matter halo --
essentially the definition of a \lq galaxy' in the present paradigm --
star clusters were known to have larger (stellar) phase space
densities than galaxies, with the dSph having the largest phase space 
densities of galaxies, but some two orders of magnitude below the lowest 
phase space densities of star clusters (\cite{walcher05}).

We here discuss the issue of what distinguishes star clusters from
galaxies, using new radial-velocity data plus new photometric surveys
and wide-field photometry of individual systems.  These data allow us to 
identify key characteristics of dark matter (see Gilmore et al.~2007   
for a comprehensive discussion). 

\section{Perspectives from New Data}

\subsection{Inferences from Photometry}

As noted above, the uniform, accurate and precise photometry from the
Sloan Digital Sky Survey (particularly the coverage of DR5) has
allowed discovery of many \lq new' satellite galaxies and star
clusters, through counts of faint stars and colour-magnitude matching.
A
very recent derivation of the satellite-galaxy luminosity function
based on the detections in DR5, correcting for the calculated
completeness and with assumed density laws for the satellite system,
is given in \cite{lf07}. No published luminosity function based on
semi-analytic prescriptions can provide a reasonable fit over the
entire satellite luminosity range.  A model including tidal stripping
(\cite{benson02}) can provide a good fit to the luminosity function of
the faintest systems, but the faint galaxies in the model have
predicted surface brightnesses many magnitudes brighter than those
of the observed systems.  This model also fails, with a shortfall of over an
order of magnitude, for the luminosity function of bright
satellites. It is clear that the models must be modified.

\begin{figure}
\centering
\includegraphics[height=3.85in, width=3in,angle=270]{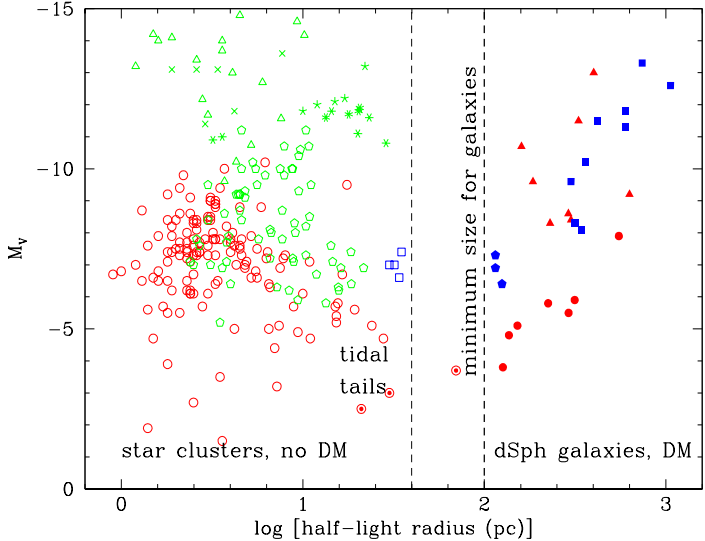}
  \caption{Absolute magnitude versus stellar half-light radius, for
  samples of star clusters (on the left of the dashed lines), including nuclear
  star clusters, young super-star-clusters, UCDs, old globular
  clusters, and  on the right of the dashed lines, dwarf spheroidal galaxies of the Milky Way and of M31. This figure is modified from Fig.~1 of Gilmore et al.~(2007), where full references are given.}
    \end{figure}

The new satellites further extend the overlap between star clusters
and galaxies in terms of total luminosity.  This is illustrated in
Fig.~1, which is a plot
of stellar half-light radius against total V-band luminosity.  The
objects represented include a comprehensive variety of star clusters,
from old galactic globular clusters through young super-star clusters,
and also nuclear stars clusters and UCDs (or Ultra Compact Dwarf,
which is probably not an apt description, discussed further below) in
galaxy clusters (see Gilmore et al.~2007 for the references).  The
known dwarf spheroidal satellite galaxies of the Milky Way (excluding
the Sagittarius dwarf) and of M31 are also represented. 

It is apparent that the characteristic stellar radius of star clusters
is less than 30pc, while the characteristic stellar radius of dSph is
greater than 100pc.  The \lq gap', delineated by the dashed lines, is occupied by one object, the
recently discovered system in Coma Berenices (Belukoruv et
al.~2007). Deep imaging with Subaru reveals an extended, irregular
structure; this, combined with the low estimated distance of $\sim
44$~kpc, suggests tides may have affected its structure.  Indeed the
globular clusters with well-defined tidal tails lie just to the left
of the \lq gap', consistent with tidal effects being important in this
size regime.  Our hypothesis is that no systems in equilibrium should
be in the \lq gap'.  Further, we interpret this figure as showing a
real division between (stable) dark-matter dominated systems
i.e.~galaxies, in which the stellar scale-length is never below 100pc,
and baryon-dominated star clusters, in which the scale-length is never
above 30pc.

In our interpretation the UCDs (the asterisks in Fig.~1) are
baryon-dominated star clusters, rather than distinct dark-matter
dominated galaxies.  This is consistent with the findings of
\cite{Hi2007} for bright UCDs in the Fornax cluster and of
\cite{Evstig07} for bright UCDs in the Virgo cluster.  The V-band
mass-to-light ratios of the UCDs are generally in the range of 3--5 in
solar units; this is higher than typical of globular clusters, but
still consistent with a purely stellar system, due to the higher
metallicity of the UCDs, typically above $-1$~dex (see these two
papers for a full discussion including comparisons with a variety of
spectral-synthesis models).  A higher mass-to-light ratio ($\sim 9$)
was inferred for a somewhat less-luminous UCD in the Virgo cluster by
\cite{has05}.  Those authors suggest that such UCDs could be the
remnant stellar nuclei of destroyed dwarf galaxies.

We  conclude that all systems with (stellar)
scale-length greater than $\sim 100$~pc have dark matter haloes, while
there are no (stable) dark-matter dominated systems with scale length
less than this.  Only pure stellar systems have very small
scale-lengths.

\subsection{Inferences from Stellar Kinematics}

Dwarf spheroidal galaxies are gas-poor, and in general the stars show
no net rotation about the centre of the dSph.  Constraints on the mass
profiles are then obtained by analyses of the stellar (random)
motions, in general the line-of-sight velocities. There are now
statistically significant samples of stars with measured line-of-sight
velocities, across the extent of the dSph on the sky -- several
hundred stars per galaxy -- for most of the `classical' dSph.  These
in general show flat or rising velocity-dispersion profiles, in
contrast to the steadily declining profiles predicted for a
mass-follows-light model -- as seen in globular clusters e.g. $\omega$~Centauri, see Fig.~3
of \cite{vdv06}.  This is compelling evidence for dark matter in the dSph.

\subsubsection{Derived Mass Profiles}

The most straightforward approach to determining the mass profile from
the line-of-sight velocities is to use the Jeans equations to analyse
the second moment of the velocity distribution in various bins in
projected radius, i.e.~the velocity dispersion profile. A full
velocity distribution-function analysis would be better, but this
requires very large samples to define the wings, and is significantly
more complicated. Where both full distribution-function
modelling and Jeans-equation modelling are available, they agree
(e.g. compare \cite{wu07} and Gilmore et al.~2007).  However, the
analysis of the stellar motions is complicated by the fact that,
without proper motions, we measure only one component of the
velocities. There is therefore a degeneracy between velocity-dispersion anisotropy and mass (see \cite{bm82} for an early
discussion).  

The simplest assumption is that the velocity-dispersion
tensor is isotropic. The derived mass profiles from Jeans-equation 
modelling, with assumed
isotropic velocity-dispersion tensor, is shown in Fig.~2, taken from
Gilmore et al.~(2007). Cold-dark-matter-dominated N-body simulations
predict a cusp at the central regions; a recent high-resolution
simulation finds that the mass density profile has a slope of $-1.2$
at 1\% of the virial radius, and asymptotes to a slope of $-1$
(\cite{diemand05}).  This prediction is also shown in Fig.~2, and
is clearly steeper than the derived profiles, which instead show a core. 
The model fits give a core radius of the mass distribution somehat
larger than that of the light, as expected since gaseous baryons dissipate to form stars.

\begin{figure}
\centering
\includegraphics[height=3in, width=3.85in]{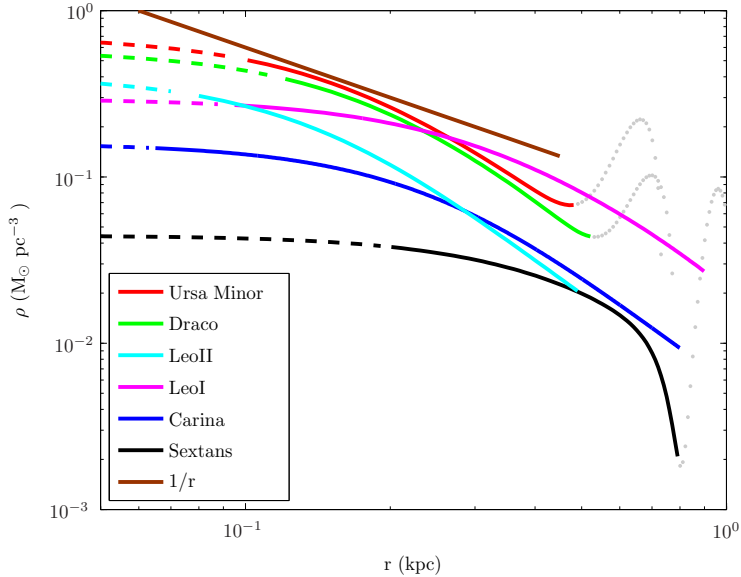}
  \caption{Derived mass density profiles from (isotropic) Jeans-equation
analyses of the stellar velocity dispersion profiles of 6 dwarf
spheroidal galaxies.  Also shown is a $r^{-1}$ profile, to which
$CDM$-mass profiles are predicted to asymptote. The modelling in each
case is reliable out to $r \simlt 500$~pc. This figure is taken from
Gilmore et al.~(2007)}
    \end{figure}

However, adopting an anisotropic velocity-dispersion tensor allows
CDM-cusps to be fitted to the same data (\cite{koch07}). 
Happily, the mass-anisotropy can be broken
for two of the dSph, using complementary independent physical
arguments, and cored mass profiles are strongly favoured.  Occam's
razor then argues that cored mass profiles are favoured for {\it
all\/} dSph. In the case of the UMi dSph, the persistence of an
observed cold sub-system is not compatible with a cusp, since the
strong gradients of the cusp would lead to disruption of the sub-system
(\cite{kleyna03}). In the case of the Fornax dSph, its globular
cluster system would have been expected to have long ago spiralled
into the galaxy center, by dynamical friction, if the mass density
profile were cusped at the centre, while a cored mass profile allows
for survival of the globulars (\cite{xavier98}, \cite{goerdt06}).

It is difficult to extend this analysis to the lowest luminosity
systems, since there simply are very few stars accessible with 8m-10m
class telescopes.

The conclusion is then that cored mass density profiles are preferred.
It is also interesting that there is a fairly narrow range of derived
mean mass densities, and that this typical value is rather low.

\subsubsection{Integrated Masses}

Mateo (1998) synthesised early results by estimating `total' masses
for dSph from their central velocity dispersions and half-light radii.
He found that a constant mass ($\sim 10^7$~M$_\odot$) was a reasonable
description of the data, albeit with a large scatter.  It is now
possible to improve on his analysis in two ways: first, by estimating
masses through integration of the mass profiles from velocity data
beyond the central regions of the galaxies, and second, by extending,
down another 3 magnitudes, the luminosity range over which mass
estimates can be made from central dispersions.  The result is seen in
Figure~3 (taken from Gilmore et al.~2007).  Here the curve of constant
mass represents $4 \times 10^7$M$_\odot$, and the scatter is
remarkably small. There is an apparently characteristic (lower) mass
to the dark matter haloes that host galaxies.

\begin{figure}[h]
\centering
\includegraphics[height=2.5in, width=3.2in]{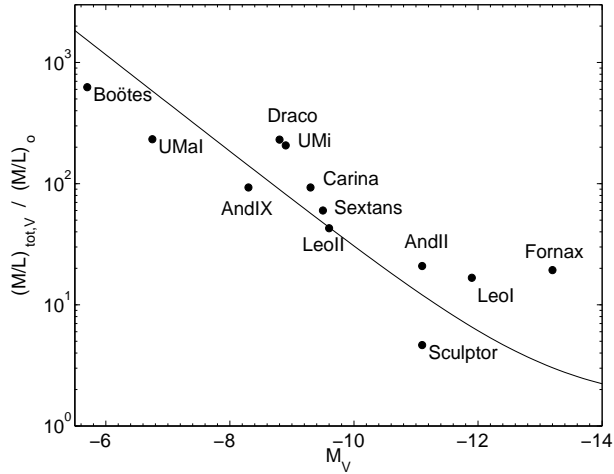}
  \caption{Mass-to-light ratios from the enclosed mass within the limit of stellar kinematic data for the profiles of Fig.~2, supplemented with galaxies for which more limited kinematic data are available, against absolute magnitude.  There is a remarkably small scatter about a line of constant mass, given by the smooth curve.  This figure is taken from
Gilmore et al.~(2007)}
    \end{figure}

There are hints from even lower luminosity galaxies, for which only
central velocity dispersions are available, that their masses may be lower
(\cite{sg07}) but the member stars are so few that the velocity dispersions
are poorly defined, and any exclusion of `outliers' of course reduces
the dispersion and reduces the derived mass.

\section{Stellar Mass Function}

It should be noted that while the (high) dark-matter content of dSph galaxies
contrasts with that of globular clusters (none), the stellar component
of the dSphs is consistent with a stellar Initial Mass Function
indistinguishable from that of the globular clusters.  The case of the
dSph in Ursa Minor is the most straightforward to analyse, since its
stellar population is old and metal-poor, of similar age and
metallicity to the halo globular clusters of the Milky Way.  Direct
star counts with the Hubble Space Telescope demonstrate no differences
between the low-mass stellar luminosity function of the UMi dSph and
that of M92 or M15, two globular clusters (see Fig.~4, taken from
\cite{wyse02}).  The initial mass function of massive stars may be
constrained by the nucleosynthetic signature that persists in the
elemental abundances of the long-lived, low-mass stars they enriched
(e.g.~review of \cite{wyse98}).  The elemental abundance data for
stars in dSph are consistent with a normal massive-star mass function,
plus the (usually extended) star formation history inferred from the
colour-magnitude diagram (e.g.~\cite{tolstoy03}; \cite{koch07b}).  This `normal' IMF is
of course equal to that seen in the solar neighbourhood and inferred
for the galactic bulge: the stellar IMF is remarkably robust,
apparently invariant over 12Gyr and a wide range of metallicities and
epochs.

Thus the stellar masses of dSph and globular clusters are quite
comparable; their dark matter contents -- and stellar scale lengths -- are
what distinguishes them.

\begin{figure}
\centering
\subfigure[]
{
\includegraphics[width=4cm,angle=270]{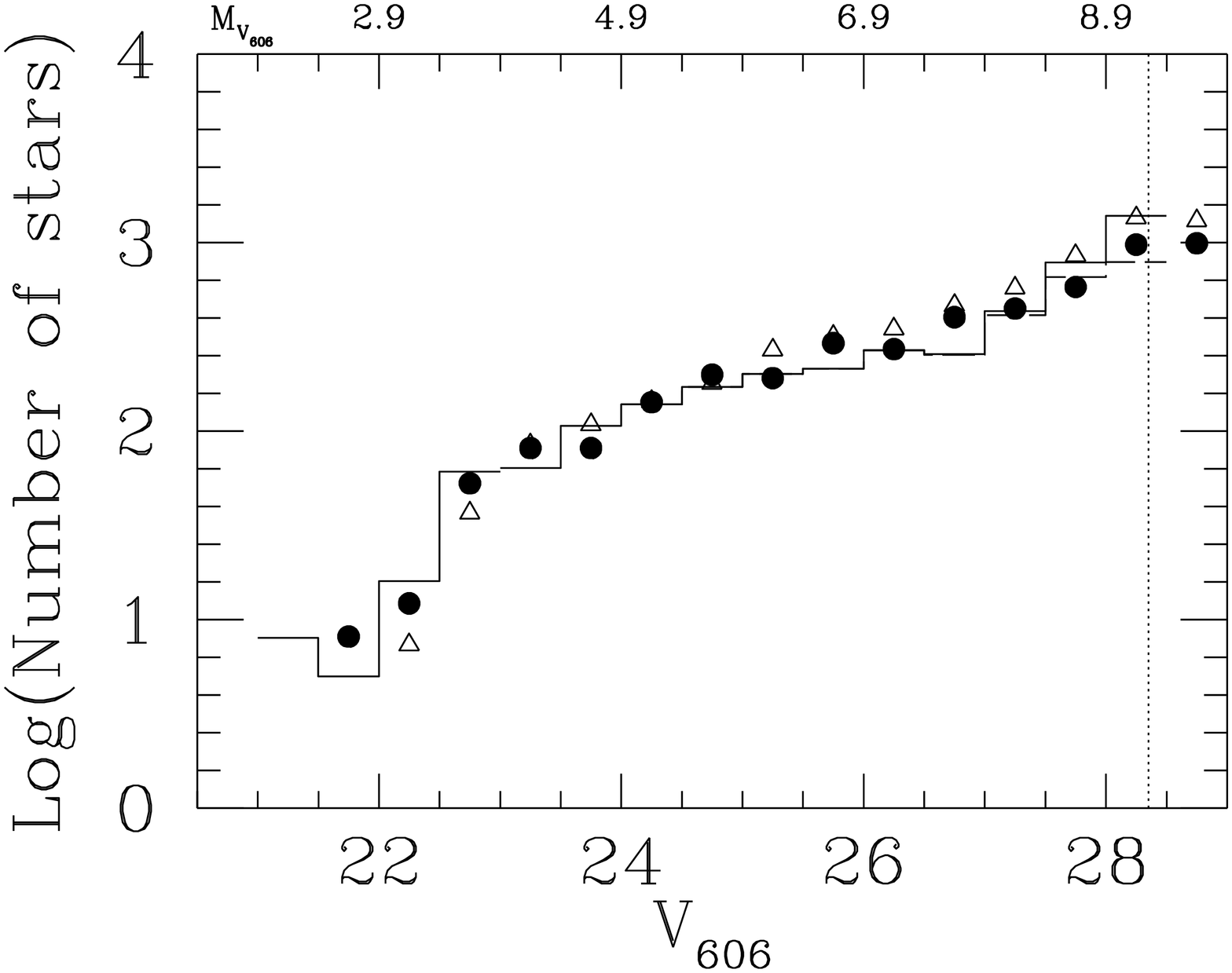}
}
\hspace{1cm}
\subfigure[]
{
\includegraphics[width=4cm,angle=270]{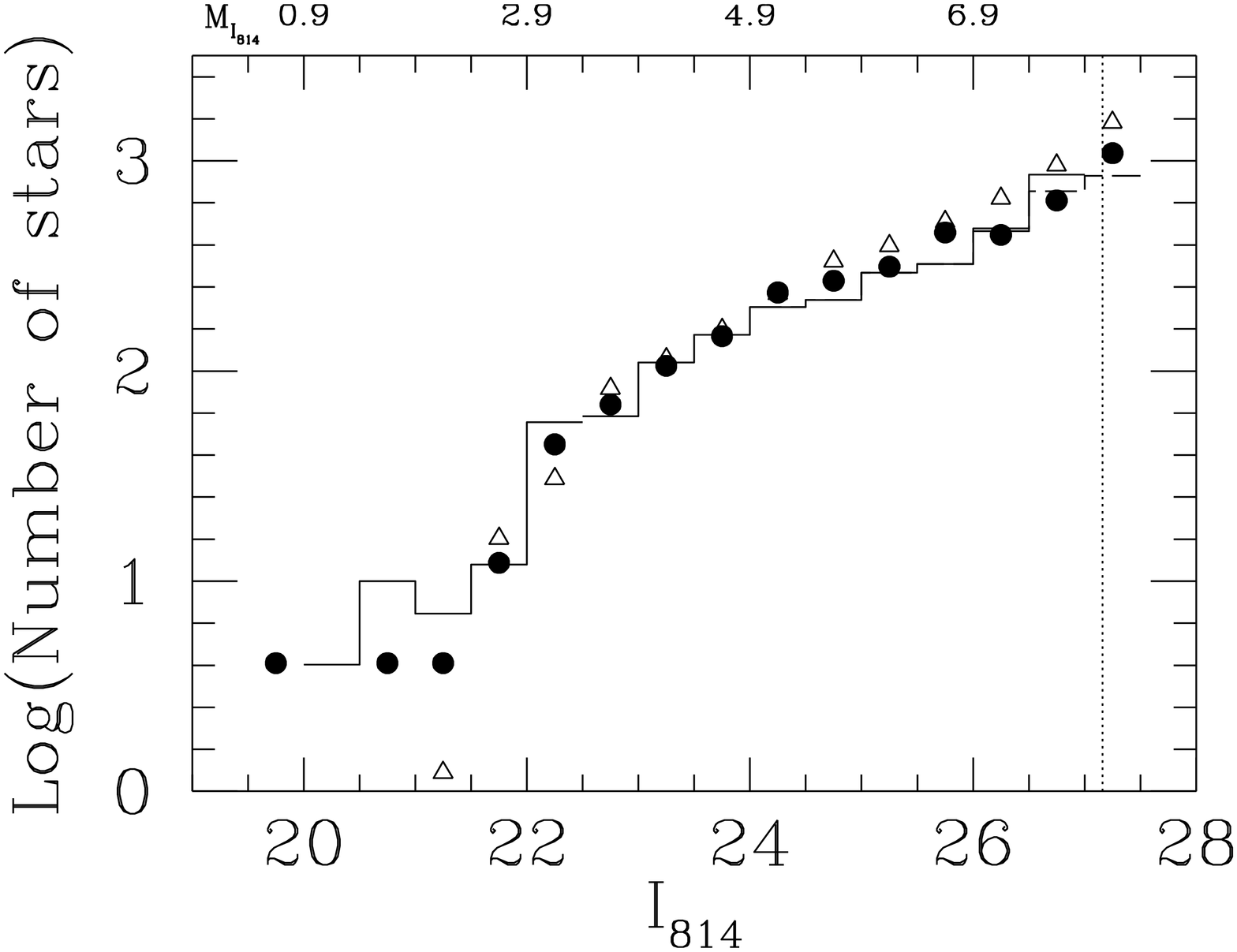}
}
\caption{Based on figures in Wyse et al.~(2002). Comparisons between the completeness-corrected Ursa Minor
luminosity functions (histograms; 
50\% completeness indicated by the vertical dotted line) 
in the V-band (a) and the I-band (b) and the same 
for M92 (filled circles) and M15 (open triangles) (both taken from
\cite{p97},  renormalized and shifted to the same distance as the Ursa Minor dSph). 
The luminosity functions for the 
globular clusters and the dwarf spheroidal galaxy are indistinguishable.}
\end{figure}

\section{Concluding Remarks}

There appears to be a minimum intrinsic scale length -- of greater
than 100~pc -- of galaxies and their associated dark matter haloes.
The density profiles of the dark matter haloes are cored, not cusped,
and have a low mean mass density of around 0.1~M$_\odot$~pc$^{-3}$, or $\sim 5~$GeV/cc, 
only around a factor of ten higher than the local dark matter density
around the position of the Sun. The combination of characteristic
scale with characteristic density leads to the expectation of a
constant mass, and this is indeed what is found.

Thus if the dark matter particle is massive -- and candidates more
massive than 100~GeV are serious candidates -- then it must be
extremely dilute to provide the central density cores. The
characteristic length scale and mass are suggestive of a
characteristic scale in the primordial power spectrum; the suppression
of small-scale power that this would imply could perhaps naturally
solve the well-known `missing satellite' problem in CDM models, together with removing the
prediction of central cusps.  Non-CDM candidates (e.g. sterile
neutrinos) need to considered seriously.

The possible effects of astrophysics in setting the stellar content of
dSph cannot be ignored in interpreting the results presented here.  Models that appeal to internal and/or external `feedback' must derive the characteristic mass self-consistently, and fit the surface brightnesses, chemical elemental abundances, star formation histories and luminosity function.  Happily there are now good observational constraints on all these aspects.  The field stellar halo and thick disk must also be included in the models, since stellar debris from disrupted satellite galaxies will most likely be deposited there. The old age of the stars in these components contrasts with the typical intermediate-age population of surviving satellites and is a major constraint (\cite{uwg96}). 

In the near future, we will have improved stellar kinematic data for
the Carina dSph (new approved VLT programme; PI: Gilmore) to which to apply a full
velocity distribution-function analysis. Improved proper motions for the dSph -- and new proper motions for the more recently discovered systems -- are needed to constrain their orbits and understand possible environmental effects on their evolution (HST application, then GAIA). The field is moving quickly. 
 
\begin{acknowledgments}

RFGW acknowledges the Aspen Center for Physics, where
this paper was written, for a stimulating environment. We thank Jon Davies for his flexibility faced with our substitution
as speakers.  We also acknowledge our collaborators in our dSph project: 
Mark Wilkinson, Jan Kleyna, Andreas Koch, Wyn Evans  and Eva Grebel.

 \end{acknowledgments}

\end{document}